\documentclass{article}
\usepackage[dvipdfmx, final]{graphicx}
\title{Diffusion of Confidential Information on Networks}
\author{Yuya Dan\\Matsuyama University}
\begin{document}
\maketitle

\section{Introduction}

The development of information and network technology enables us to share information at anywhere and anytime.
The social networks on the Internet,
for example BBS, SNS, Twitter etc. in the Web,
have become the personal network media of information.
They are quite useful in information sharing each other.

On the other hand,
we faced the difficulties in the confidentiality of sensitive content,
which have the risk of diffusion of personal or secret information over the Internet.
Once the personal and secret information diffuses on the Internet,
people on the Internet can know the information anywhere and anytime even in future.

This paper presents mathematical models for diffusion phenomena of personal or secret information on the Internet in particular.
We investigate the behavior of the models using analytical and computational methods with numerical Monte Carlo simulation.
We also consider the structure and dynamics of diffusion on networks constructed by a large number of people with interaction or communication each other.

As is well known that social networks have grown rapidly on the Internet.
The community on the Internet is in general visible from access logs in servers,
rather than that in the real world,
so that we can easily analyze the structure and dynamics of the social networks.
Not only the element but also the link of a social network determines the behavior of social systems.

A network \cite{Newman} is a set of points (also called \textit{vertices} or \textit{nodes}) connected by lines (also called \textit{edges} or \textit{links}).
We may call \textit{complex networks},
if the number of points and links is so large that only computational calculation can analyze them.
Any system with coupled elements can be represented as a network,
so that our world is full of networks\cite{Dorogovtsev}.

This is a natural generalization of the previous work by Dan \cite{Dan}.
In this paper,
we consider the diffusion phenomena of personal or secret information on the variety of networks,
such as complete, random, stochastic and scale-free networks.

\section{Network Models}

In this section,
we provide each definition of the corresponding models of networks.
The dynamics of diffusion or percolation depends on the structure of networks.
We see the property of networks under the definition,
and consider the characteristics of each network.

\subsection{Complete Networks}

A complete network is the network all of whose two vertices have an edge.
There is no pair that does not have edge in the network.
When the number of vertices is $n$,
the network has $n ( n - 1 ) / 2$ edges.

As the previous work,
Dan \cite{Dan} investigaed the mathematical modeling and computer simulation of diffusion phenomena on social networks for complete networks.

\subsection{Random Networks}

A random network is the network whose vertices have edges at random.
Randomness is assumed for not only uniform distribution,
but also any possible function of distribution.
In this paper,
we assume uniform distribution of randomness.

\subsection{Stochastic Networks}

A stochastic network is the network whose each edge has probability of the value between zero and one.
Each edge mediates the information at the probability that depends on the edge.
One can communicate on the edge at the probability $p$,
on the other hand,
one cannot communicate on the edge at the probability $1 - p$.
The possibility of communication depends on the probability $p$ defined each on the edge.

\subsection{Scale-free Networks}

A scale-free network is defined the power law for the number of edges.
There are some vertices,
which are called \textit{hubs},
that have comperable large number of edges.
On the other hand,
almost all vertices have only a few edges.
The graph of the number of edges indicates the law of power.
Scale-free networks are first proposed as small-world networks by Watts and Strogatz \cite{Watts-Strogatz}.

It is known that scale-free networks have high cluster coefficients like regular lattices.
However, these networks have small characteristic path lengths like random networks.

\section{Simulation}

Let us begin with the setting of constructing the structure of the networks in the simulation.

\subsection{Comparison between random and stochastic networks}
As random networks,
we provide edges between two vertices on the network at uniform probability of $1 / 2$.
We expect that there are $n / 2$ edges at random.

Figure \ref{RandomNetworkMatrix} indicates a link matrix for the random networks we constructed.
The ( $i$, $j$ ) element of the matrix is 1 if the vertex $i$ and $j$ are connected,
0 if not.
Therefore,
this matrix is symmetric.

\begin{table}
\centering
\caption{Link matrix for random networks}
\label{RandomNetworkMatrix}
\begin{tabular}{ccccccccccccccccccc}
\hline
1&0&1&0&0&1&1&0&0&0&1&0&0&0&1&1&1&1&$\cdots$\\
0&1&0&1&0&0&1&1&1&0&0&1&0&0&0&1&1&0&$\cdots$\\
1&0&1&1&1&0&0&0&1&0&0&0&0&1&1&1&0&1&$\cdots$\\
0&1&0&1&0&0&0&0&1&0&0&1&1&0&1&1&0&0&$\cdots$\\
0&0&1&0&1&0&1&0&1&1&0&0&1&1&0&1&1&1&$\cdots$\\
\multicolumn{19}{c}{$\cdots$}\\
\hline
\end{tabular}
\end{table}

As stochastic networks,
we provide the probabilities to all of edges on the network.
The value of the probabilities take zero to one uniformly.
We expect that any edge has probability of $1 / 2$ in average.

Figure \ref{StochasticNetworkMatrix} indicate a probability matrix for stochastic networks.
This matrix is also symmetric.
All of edges in the network has vakues between 0 and 1,
although the value of diagonal elements have no sense.

\begin{table}
\centering
\caption{Probability matrix for stochastic networks}
\label{StochasticNetworkMatrix}
\begin{tabular}{cccccccccccc}
\hline
0.00&0.56&0.19&0.81&0.59&0.48&0.35&0.90&0.82&0.75&$\cdots$\\
0.56&0.15&0.95&0.14&0.91&0.69&0.30&0.43&0.07&0.97&$\cdots$\\
0.19&0.95&0.29&0.44&0.23&0.58&0.53&0.63&0.16&0.50&$\cdots$\\
0.81&0.14&0.44&0.07&0.78&0.52&0.61&0.96&0.07&0.88&$\cdots$\\
0.59&0.91&0.23&0.78&0.36&0.86&0.23&0.86&0.23&0.25&$\cdots$\\
\multicolumn{11}{c}{$\cdots$}\\
\hline
\end{tabular}
\end{table}

Despite two matrices have different elements,
the averages of all elements are expected same as $1 / 2$.
That is,
\begin{equation}
	\lim_{n \to \infty} \frac{1}{n^2} \sum_{i, j} r_{ij}
	= \lim_{n \to \infty} \frac{1}{n^2} \sum_{i, j} p_{ij}
	= \frac{1}{2}
\end{equation}
where $r_{ij}$ are the elements of the link matrix of random networks and $p_{ij}$ are the elements of the probability matrix of stochastic networks.

We investigate the increase of the number of people who knows personal or secret information.

\subsection{Diffusion on scale-free networks}

In order to use scale-free network,
we have to construct one before that.
In fact,
we can construct the scale-free networks according to the growth model \cite{Newman-Barabasi-Watts} of networks as follows.

First of all,
we provide one vertex on the empty network, which may not be a network.
Next,
providing a vertex with an edge which connects to a vertex already exists at the rate of probability proportional to the number of edges.
This procedure accelerates the growth of hubs and long tail structure.
Continuing the procedure of making a network until the number of vertex we expect in the simulation,
we can finally obtain the network with scale-free structure.

In general,
it is easy to find the structure in the real world.
We know that the networks of friendship, the Internet, Web pages, SNS users, Twitter and so on forms scale-free,
so that we can see hubs in the network and the typical structure of scale-free networks.
In our discussion,
personal or secret information are important in the digital societies.
Once these confidential informaion are known widely on the networks,
we cannot recover that situation.

\begin{figure}
\centering
\rotatebox{0}{\includegraphics[width=8.4cm, clip]{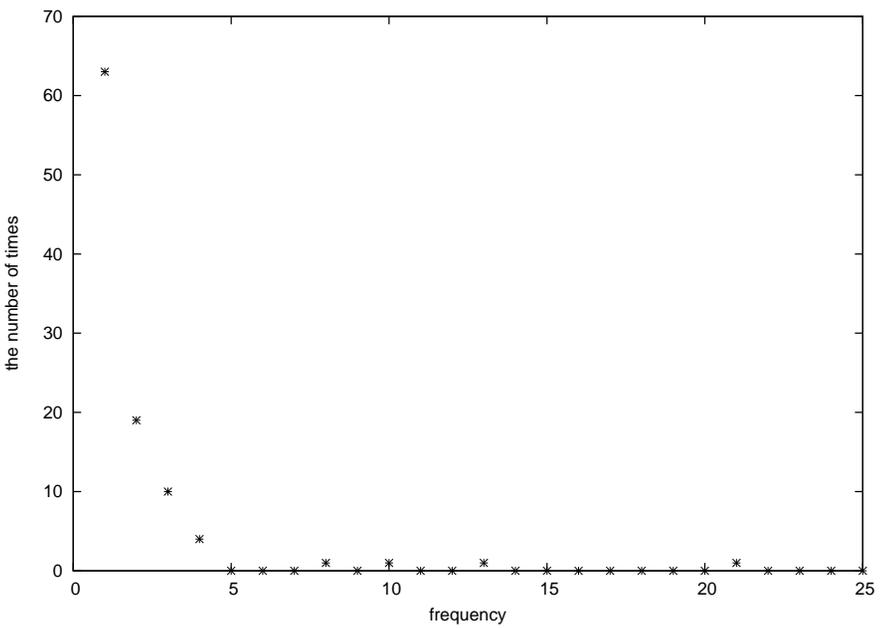}}
\caption{Graph of the law of power on scale-free networks}
\label{PowerLaw}
\end{figure}

It is Figure \ref{PowerLaw} that we have constructed in the settings of simulation.
The vertex which has 21 edges is a hub in the network.
On the other hand,
over 60\% of vertices have only one edge which is at minimum.
This graph shows the power law approximately.

\section{Results}

\begin{figure}
\centering
\rotatebox{0}{\includegraphics[width=8.4cm, clip]{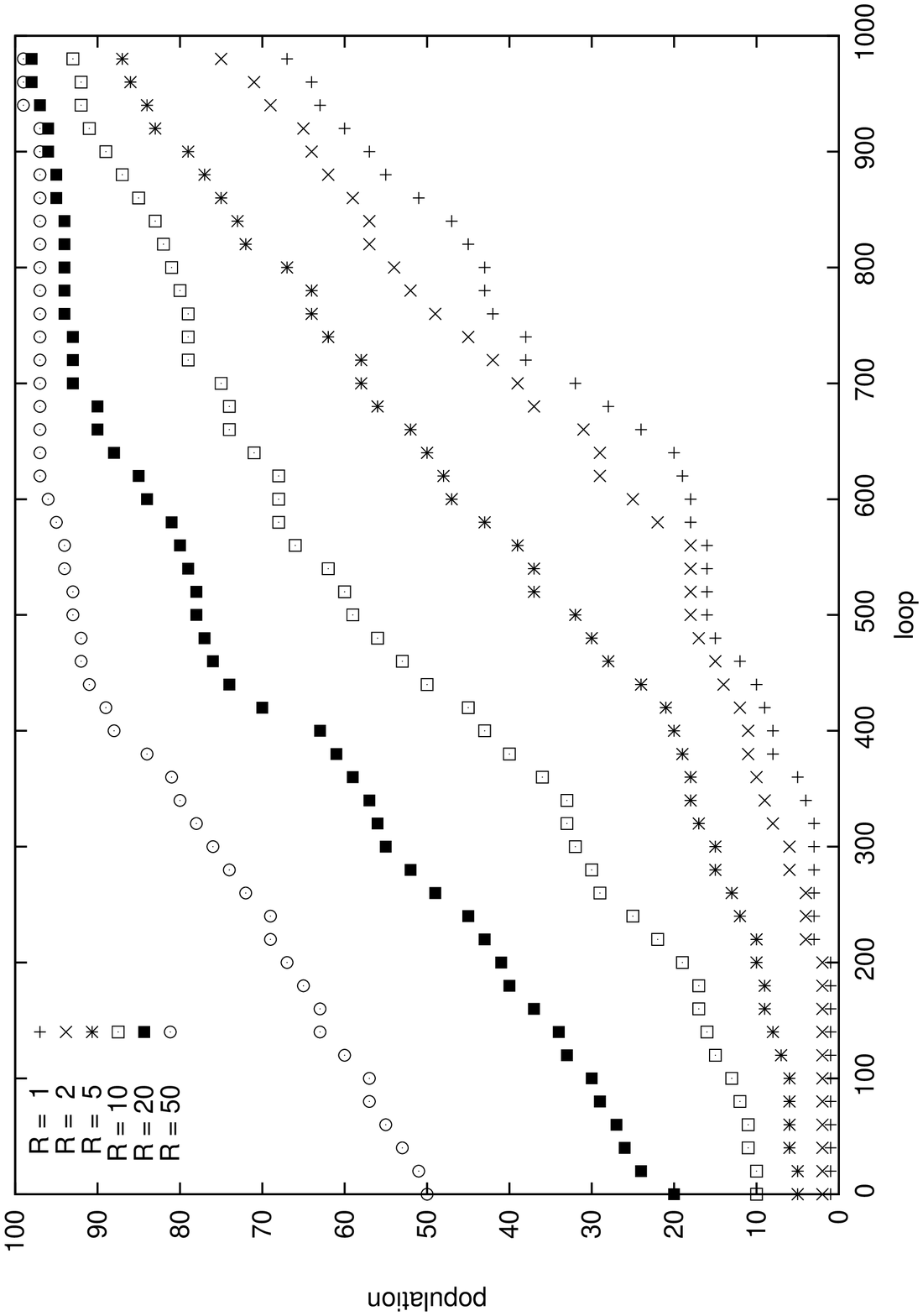}}
\caption{Graph of the time sequential diffusion on random networks}
\label{RandomNetwork}
\end{figure}

\begin{figure}
\centering
\rotatebox{0}{\includegraphics[width=8.4cm, clip]{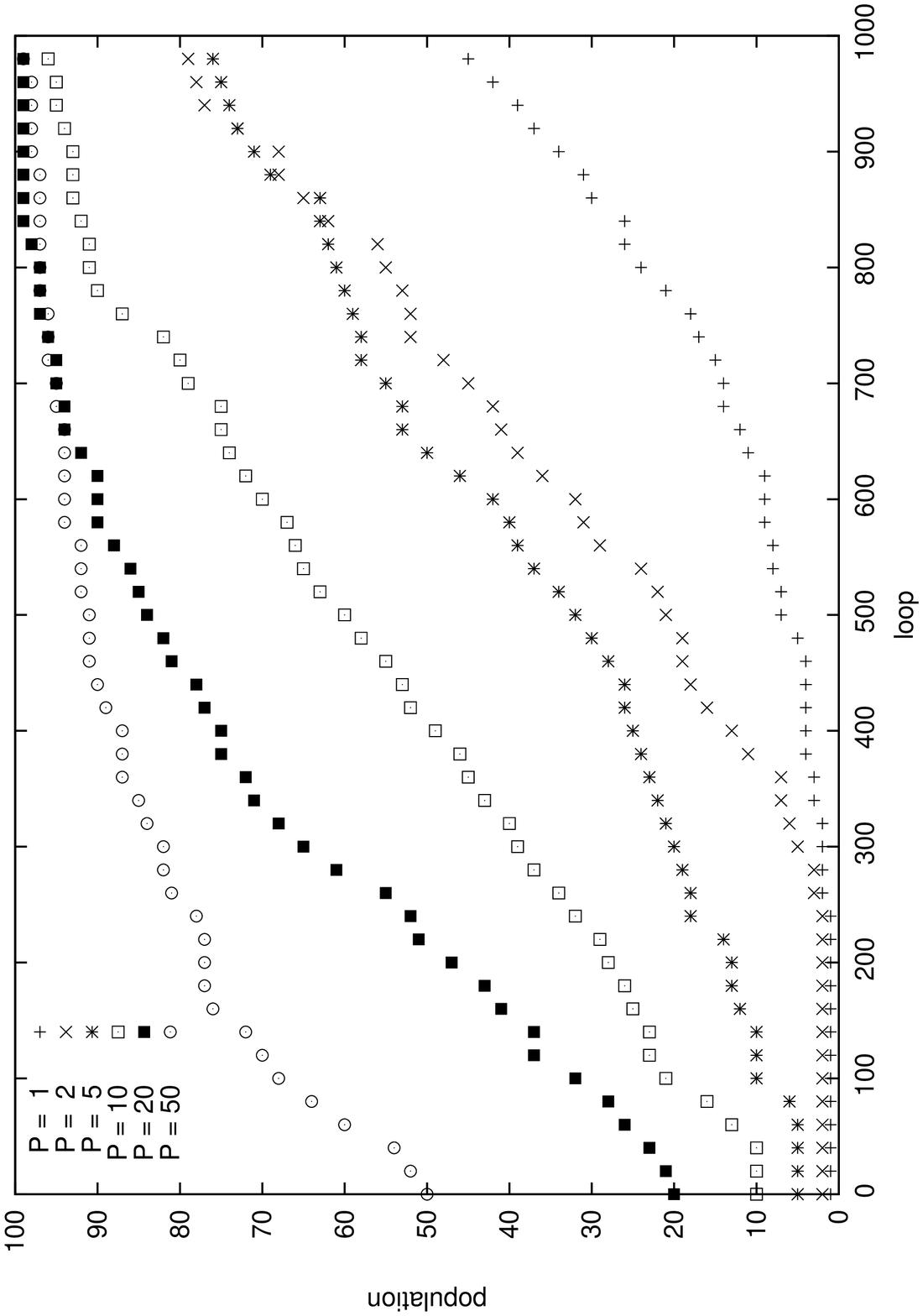}}
\caption{Graph of the time sequential diffusion on stochastic networks}
\label{StochasticNetwork}
\end{figure}

As the result of the Monte Carlo simulation with initial value of 1, 2, 5, 10, 20 and 50 of maximum polulation 100,
we obtain the diffusion phenomena both of random networks (See Figure \ref{RandomNetwork}) and of stochastic networks (See Figure \ref{StochasticNetwork}).
The population means the number of people who knows personal or secret information.
We cannot see the significant difference between the diffusion on the stochastic networks and that on random networks.

\begin{figure}
\centering
\rotatebox{0}{\includegraphics[width=8.4cm, clip]{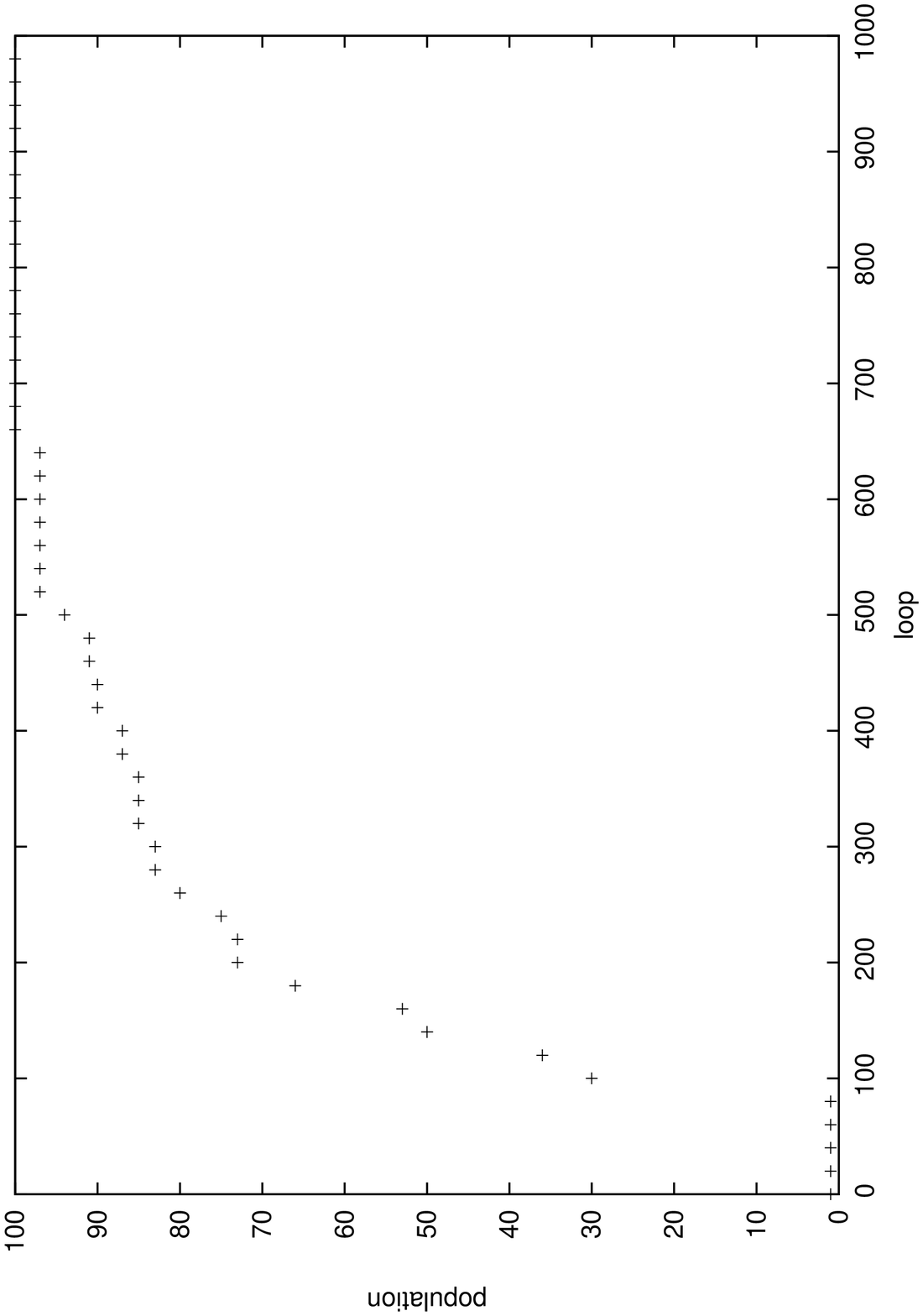}}
\caption{Graph of the time sequential diffusion on scale-free networks}
\label{ScaleFreeNetwork}
\end{figure}

Next,
we obtain the result of rapid diffusion on scale-free networks.
Figure \ref{ScaleFreeNetwork} indicates the graph of the time sequential diffusion on scale-free networks.
We can see a large quantum leap in the loop time from 80 to 100.
It seems to occur there that the hub diffused the infomation to a lot of vertices connected with the hub.

There is no doubt that saturation occurs at the first loop if the structure of the network is complete.

\section{Concluding Remarks}

We have discussed the diffusion of confidential information on a variety of networks.
It should be remarked that there is no significant difference between random networks and stochastic networks.
It becomes clear that we should afraid to use personal or secret information on scale-free networks.


\begin{thebibliography}{99}
\bibitem{AralBrynjolfssenAlstyne}
	S.~Aral, E.~Brynjolfsson, M.~W.~V.~Alstyne,
	\newblock "Productivity Effects of Information Diffusion in Networks,"
	\newblock \textit{MIT Center for Digital Business},
	\newblock  Working Paper \#234.
	\newblock (2007)
	\newblock Available at SSRN: http://ssrn.com/abstract=987499

\bibitem{ChristakisFowler}
	N.~A.~Christakis, J.~H.~Fowler,
	\newblock "The Spread of Obesity in a Large Social Network over 32 Years,"
	\newblock \textit{New England Journal of Medicine},
	\newblock 357, pp.~370--379.
	\newblock (2007)

\bibitem{Dan}
	Y.~Dan,
	\newblock "Modeling and Simulation of Diffusion Phenomena on Social Networks,"
	\newblock to appear in \textit{The proceedings of 2011 Third International Conference on Computer Modeling and Simulation}.
	\newblock (2011)

\bibitem{Dellarocas}
	C.~Dellarocas,
	\newblock "The Digitization of Word of Mouth: Promise and Challenges of Online Feedback Mechanisms,"
	\newblock \textit{Management Science},
	\newblock 49, 10, pp.~1407--1424.
	\newblock (2003)

\bibitem{Dorogovtsev}
	S.~N.~Dorogovtsev,
	\newblock \textit{Lectures on Complex Networks},
	\newblock Oxford University Press.
	\newblock (2010)

\bibitem{HuckfeldtSprague}
	R.~Huckfeldt, J.~Sprague,
	\newblock "Discussant Effect on Vote Choice: Intimacy, Structure and Interdependence,"
	\newblock \textit{The Journal of Politics},
	\newblock 53, 1, pp.~122--158.
	\newblock (1991)

\bibitem{LeskovecAdamicHuberman}
	J.~Leskovec, L.~Adamic, B.~A.~Huberman,
	\newblock "The Dynamics of Viral Marketing,"
	\newblock \textit{ACM Transactions on the Web},
	\newblock Vol.~1, Iss.~1, Article No.~5.
	\newblock (2007)

\bibitem{Newman}
	M.~E.~J.~Newman,
	\newblock \textit{Networks},
	\newblock Oxford University Press.
	\newblock (2010)

\bibitem{Newman-Barabasi-Watts}
	M.~E.~J.~Newman, A.~L.~Barab\'asi, D.~J.~Watts,
	\newblock \textit{The Structure and Dynamics of Networks},
	\newblock Princeton University Press.
	\newblock (2006)

\bibitem{Rogers}
	E.~M.~Rogers,
	\newblock \textit{Diffusion of Innovations},
	\newblock Free Press, New York.
	\newblock (1995)

\bibitem{VazquezPastor-SatorrasVespignani}
	A.~V\'azquez, R.~Pastor-Satorras, A.~Vespignani,
	\newblock "Large-scale topological and dynamical propertes of the Internet,"
	\newblock \textit{Physical Review E},
	\newblock Vol.~65, No.~066130.
	\newblock (2002)

\bibitem{Verhulst}
	P.-F.~Verhulst,
	\newblock "Notice sur la loi que la population poursuit dans son accroissement,"
	\newblock \textit{Correspondance math\'ematique et physique}
	\newblock 10, pp.~113--121.
	\newblock (1838)

\bibitem{Watts-Strogatz}
	D.~J.~Watts and S.~H.~Strogatz,
	\newblock "Collective dynamics of 'small-world' networks,"
	\newblock \textit{Nature}
	\newblock 393, pp.~440--442.
	\newblock (1998)

\end{thebibliography}
\end{document}